\documentclass[11pt,a4paper]{JHEP}
\usepackage{epsfig}
\usepackage{latexsym}
\usepackage{psfrag}
\usepackage{amssymb}

\catcode`\@=11

\def\marginnote#1{}

\def\draftlabel#1{{\@bsphack\if@filesw {\let\thepage\relax
   \xdef\@gtempa{\write\@auxout{\string
      \newlabel{#1}{{\@currentlabel}{\thepage}}}}}\@gtempa
   \if@nobreak \ifvmode\nobreak\fi\fi\fi\@esphack}
        \gdef\@eqnlabel{#1}}
\def\@eqnlabel{}
\def\@vacuum{}
\def\draftmarginnote#1{\marginpar{\raggedright\scriptxize\tt#1}}

\def\draft{\oddxidemargin -.5truein
        \def\@oddfoot{\sl preliminary draft \hfil
        \rm\thepage\hfil\sl\today\quad\militarytime}
        \let\@evenfoot\@oddfoot \overfullrule 3pt
        \let\label=\draftlabel
        \let\marginnote=\draftmarginnote
   \def\@eqnnum{(\theequation)\rlap{\kern\marginparsep\tt\@eqnlabel}%
\global\let\@eqnlabel\@vacuum}  }

\def\numberbysection{\@addtoreset{equation}{section}
        \def\theequation{\thesection.\arabic{equation}}}

\def\underline#1{\relax\ifmmode\@@underline#1\else
        $\@@underline{\hbox{#1}}$\relax\fi}

\def\titlepage{\@restonecolfalse\if@twocolumn\@restonecoltrue\onecolumn
     \else \newpage \fi \thispagestyle{empty}\c@page\z@
        \def\thefootnote{\fnsymbol{footnote}} }

\def\endtitlepage{\if@restonecol\twocolumn \else \newpage \fi
        \def\thefootnote{\arabic{footnote}}
        \setcounter{footnote}{0}}  

\catcode`@=12
\relax

\makeatletter

\def\publist{\@ifnextchar[{\@publist}{\@@publist}}
\def\@publist[#1]{\list
        {[\arabic{pubctr}]\hfill}{\settowidth\labelwidth{[999]}
        \leftmargin\labelwidth
        \advance\leftmargin\labelsep
        \@nmbrlisttrue\def\@listctr{pubctr}
        \setcounter{pubctr}{#1}\addtocounter{pubctr}{-1}}}
\def\@@publist{\list
        {[\arabic{pubctr}]\hfill}{\settowidth\labelwidth{[999]}
        \leftmargin\labelwidth
        \advance\leftmargin\labelsep
        \@nmbrlisttrue\def\@listctr{pubctr}}}
 \relax
\makeatother
\newskip\humongous \humongous=0pt plus 1000pt minus 1000pt

\newif\ifdtup

\relax

\newcommand{\be}{\begin{equation}}
\newcommand{\ee}{\end{equation}}
\newcommand{\bq}{\begin{eqnarray}}
\newcommand{\eq}{\end{eqnarray}}
\newcommand{\bqs}{\begin{eqnarray*}}
\newcommand{\eqs}{\end{eqnarray*}}
\newcommand{\Hil}{\mathcal{H}}
\newcommand{\vac}{|0\rangle}
\newcommand{\leftvac}{\langle 0 |}
\newcommand{\state}[1]{| #1 \rangle}
\newcommand{\etats}[1]{\langle #1 |}
\newcommand{\p}{\partial}
\newcommand{\parder}[1]{\frac{\partial}{\partial #1}}

\newcommand{\IZ}{\mathbb{Z}}

\newcommand{\crea}{a^\dagger}

\newcommand{\erf}{\mbox{erf}}

\newcommand{\refpj}[1]{~(\ref{#1})}
\newcommand{\sfi}{string field}
\newcommand{\sft}{string field theory}

\newcommand{\half}{\frac{1}{2}}
\newcommand{\bea}{\begin{eqnarray}}
\newcommand{\eea}{\end{eqnarray}}

\def\l{\lambda}

\def\G{\Gamma}

\title{Toy Model for Tachyon Condensation in Bosonic String Field Theory}
\author{ Pieter-Jan De Smet and Joris Raeymaekers
\\Instituut voor theoretische fysica, Katholieke
Universiteit Leuven,\\
Celestijnenlaan 200D, B-3001 Leuven, Belgium. \\
E-mail: {\tt Joris.Raeymaekers, Pieter-Jan.DeSmet@fys.kuleuven.ac.be} }
\abstract{
We study tachyon condensation in a baby version of Witten's open string field
theory. For some special values of one of the parameters of the model, we are
able to obtain closed form expressions for the stable vacuum state and for 
the value of the potential at the minimum. We study the convergence rate of the
level truncation method and compare our exact results with the numerical results
found in the full string field theory.} 
\preprint{ KUL-TF-2001/17 \\
  {\tt hep-th/0107070}}

\begin{document}

In this letter, we discuss a simple toy model for tachyon
condensation in bosonic string field theory. The full string
field theory   problem \cite{kost}--\cite{mt}
consists of extremising a complicated
functional on the Fock space built up from an infinite number of
matter and ghost oscillators. As a first simplification, one can
consider the variational problem in the restricted Hilbert space
of states generated by a single matter oscillator. This problem
is still rather nontrivial because the restricted Hilbert space
still contains an infinite number of states. The model we will
consider here is precisely of this form and
 its behaviour closely resembles the one found in the full theory
with level approximation methods. The main simplification
lies in the limited number of degrees of freedom and the fact
that we don't have to deal with the technicalities of the ghost system.

The motivation for considering 
such simplified models is twofold.
First of all, the level approximation method to the full string theory
problem remains largely `experimental': there doesn't seem to be a
convincing a priori reason why this approximation scheme  converges to
the exact answer, nor do we have any information about the rate of convergence
except the `experimental' information we have from considering the
first few levels. Our toy model will allow for the derivation of
 exact results on the convergence of the level truncation method
albeit in  a not fully realistic context.

The second reason for considering toy models is perhaps more fundamental:
 it would be of considerable
interest to obtain the exact solution for the stable vacuum in the full
theory. 
Such an exact solution would allow  a detailed description of the physics 
around the stable vacuum, where interesting phenomena expected
to arise \cite{closedstrings}.
However, despite many efforts, 
this solution is lacking at the present
time\footnote{See e.g.~\cite{kostpotting2} where a recursive technique was
formulated. Other exact solutions are known, see for example~\cite{Tak}.}. 
The model we will consider is in some sense the
`minimal' problem one should be able to solve if one hopes to find an
analytic solution to the full problem\footnote{Other toy models for tachyon condensation 
were considered in~\cite{Minahan,padic,GP}.}.

\paragraph{}
In section~\ref{TM:action} we will give the action of the toy model. 
In its most general form, the model depends  on some parameters that 
enter in the definition of a  star product and are the analogue of 
the Neumann coefficients in bosonic \sft. These parameters
are further constrained if we insist that the toy model star product
satisfies some of the properties that are present in the full \sft.
More specifically, the \sft\ star product satisfies the following properties:
\begin{itemize}
\item
The three-string interaction term is cyclically symmetric.
\item
The star product is associative.
\item
Operators of the form $a - a^\dagger$ act as derivations of the star-algebra.  
\end{itemize}

We impose cyclicity of the interaction term in our toy 
model in
section~\ref{TM:cyc}. We deduce the equations of 
motion in section~\ref{TM:eom}. 
In section~\ref{TM:eom} we define 
the star product for the toy model. 
We discuss the restrictions following from imposing associativity of
the star product in section~\ref{TM:ass}. It turns out
that we are left with 3
different possibilities, hereafter called case I, II and
III. As is the case for the bosonic \sft\ we can also look if there is a
derivation $D = a - a^\dagger$ of the star-algebra. This further restricts the
cases I, II and III to case Id, IId and again Id respectively. This is 
explained
in section~\ref{TM:der}, where we also discuss the existence of an identity
of the star-algebra.

After having set the stage we can start looking for exact solutions. In 
section~\ref{TM:CaseI} we give the exact results for case I. In particular we
are able to write down closed form expressions for the stable vacuum, the
effective potential and its branch structure and the convergence rate of the 
level truncation method. 
We also compare these results with the behaviour found in bosonic
\sft.
In section~\ref{TM:othersols} 
we mention the other exact solutions we have found.
In  section~\ref{TM:CaseIId}, we discuss the case IId which perhaps bears
the most resemblance to the full string field theory problem.
In this case, it is possible  to recast the equation of motion in
the form of an ordinary second order nonlinear differential equation.
This equation is not of the Painlev\'{e} type and we have not been able to
find an exact solution. 
Here too, it is possible to get very accurate information about the
stable vacuum using the level truncation method.
We conclude in
section~\ref{TM:end} with some suggestions for further research. 
\section{The action}\label{TM:action}	
The toy model we consider has the following action 
(the potential energy is equal to minus the action):
\be\label{Toyaction}
S(\psi) = - \frac{1}{2} \etats{\psi}(L_0-1) \state{\psi} -
 \frac{1}{3} \etats{V}
\state{\psi} \state{\psi}\state{\psi}
\ee 
where $L_0$ is the usual kinetic operator $L_0= a^\dagger a$ 
and $[a,a^\dagger] = 1$. 
 Let us denote the Fock space which is built up in the
usual way by $\Hil$. 
The ``string field'' $\state{\psi}$ is simply a state in this
Fock space $\Hil$ and can thus be expanded as 
$$
\state{\psi} = \psi_0 \vac + \psi_1 a^\dagger\vac + \psi_2 (a^\dagger)^2\vac +
\cdots,
$$
where the coefficients $\psi_0, \psi_1, \cdots$ 
are complex numbers. To illustrate the analogy
between this toy model and Witten's bosonic \sft\ \cite{wittenbsft}, 
the complex numbers
$\psi_i$ in the toy model correspond to the space-time fields in the bosonic
\sft\ in the Siegel gauge. The term $-1$ in the kinetic part of the action 
should be thought of as the zero point energy in the bosonic string. 
In this way, the state $\vac$ has negative energy. 

The interaction term is defined  as follows:
\be\label{Toyint} 
\etats{V}
\state{\psi} \state{\psi}\state{\psi}=\  
_{123}\leftvac \exp ( \frac{1}{2} \sum_{i,j=1}^3 N_{ij} a_i a_j)\
\state{\psi}_1\state{\psi}_2\state{\psi}_3.
\ee
The numbers $N_{ij}$ mimic the Neumann coefficient in Witten's
\sft~\cite{GrossJev}. 
There they carry additional indices $N_{ij,kl}\ \eta^{\mu\nu}$ where $k,l = 1,
\ldots,\infty$ label the different modes of the string and $\mu,\nu =
1,\ldots,26$
are space-time indices.

We have introduced 
 three copies of the Fock space $\Hil$. The extra
subscript on a state denotes the copy the state is in:
\bqs
&&\mbox{if } \state{\psi} = \sum_m \psi_m a^{\dagger m} \vac\ \in \Hil, \\
&&\mbox{then } 
\state{\psi}_i = \sum_m \psi_m a^{\dagger m}_i \vac_i \in \Hil_i\quad\mbox{  
for }
i=1,2,3.
\eqs
By definition we have the following commutation relations in
$\Hil_1\otimes\Hil_2\otimes\Hil_3$:
\be\label{Toycomm}
[a_i,a_j^\dagger] = \delta_{ij}.
\ee
Hence the interaction term\refpj{Toyint} of the toy model is the inner product
between the state $\state{V} = \exp ( \frac{1}{2} \sum N_{ij}
a_i^\dagger a_j^\dagger)\vac_{123} \in\Hil_1\otimes\Hil_2\otimes\Hil_3$ and 
$\state{\psi}_1 \otimes\state{\psi}_2 \otimes \state{\psi}_3$.

As an example let us calculate the action for 
$\state{\psi} = t \vac + u\ a^\dagger \vac,$ i.e. the level 1 part of the 
action\refpj{Toyaction}.
The kinetic part is obviously
$$-{1 \over 2} t^2 $$
and the interaction is
\bqs
&&{1 \over 3}\ _{123}\leftvac \left(1+ {1 \over 2} N_{ij}a_ia_j\right) 
\left(t + u\ a_1^\dagger\right) \left(t + u\ a_2^\dagger\right) 
\left(t + u\ a_3^\dagger\right)
\vac_{123} = \\
&&{1 \over 3} t^3 + {1 \over 3}\left( N_{12}+N_{13}+N_{23}\right)t u^2
\eqs
\section{Cyclicity}\label{TM:cyc}
As is the case for the full \sft\ we would like the interaction to be 
\emph{cyclic}\index{cyclicity!in toy model}:
\be\label{Toycyc} 
\etats{V} \state{A}\state{B}\state{C} = \etats{V} \state{B}\state{C}\state{A}.
\ee
Let us see what restrictions this gives for the matrix $N$. Imposing the
cyclicity\refpj{Toycyc} leads to $N_{11} = N_{22} = N_{33}$,  
$N_{12} = N_{23} = N_{31}$ and $N_{13} = N_{21} = N_{32}$. Hence $N$ will be of
the following form:
$$ N = \left(\begin{array}{ccc}
N_{11} &N_{12} &N_{13}\\
N_{13}& N_{11} & N_{12}\\
N_{12}& N_{13}  & N_{11}\\
\end{array} \right)
.$$
Because the oscillators $a_1,a_2$ and $a_3$ commute among each other, the matrix
$N$ can be chosen to be symmetric without losing generality. Hence we have 
fixed  the matrix $N$ to be of the form
$$ N_{ij} = \left(\begin{array}{ccc}
2 \lambda &\mu &\mu\\
\mu&2 \lambda & \mu\\
\mu& \mu  &2 \lambda\\
\end{array} \right)
.$$
From this form it is clear that imposing cyclicity in our toy-model forces 
the star product to be commutative as well.
\section{The equation of motion}\label{TM:eom}
If we impose the condition that the interaction is cyclically symmetric, 
the equation of motion reads
\be\label{Toyeom}
(a^\dagger a -1) \state{\psi} +\state{\psi} \ast \state{\psi} = 0. 
\ee
Here we have introduced the star product, it is defined by
\bq\label{Toystar}\index{star product!in toy model}
&&\state{\psi} \ast \state{\eta} =\\
&&_{23}\leftvac \exp( \frac{1}{2} \sum_{i,j = 2}^3 N_{ij} a_i a_j +
\sum_{i=2}^3 a_1^\dagger N_{1i} a_i + 
\frac{1}{2} a_1^\dagger N_{11} a_1^\dagger )\
\vac_1 \state{\psi}_2\state{\eta}_3\nonumber
\eq
Let us give some examples of the star product:
$$\vac \ast \vac = e^{\lambda a^{\dagger 2}}\vac.$$
The star product of two coherent states gives a squeezed state
\bqs
&&e^{l_1 \crea} \vac \ast e^{l_2 \crea} \vac =\\
&& \exp\left( \lambda (l_1^2 + l_2^2) + \mu\ l_1 l_2 \right)
\exp \left( \lambda a^{\dagger 2} + \mu ( l_1 + l_2) \crea\right) \vac
.\eqs
By taking derivatives one can calculate lots of star products e.g.
\bqs
\vac\ast\crea\vac &=& \mu\crea e^{\lambda a^{\dagger 2}}\vac\\
\crea\vac\ast\crea\vac &=& (\mu+\mu^2 a^{\dagger 2}) e^{\lambda a^{\dagger 2}}\vac\\
\eqs
We will now  write the equations (\ref{Toyeom})
 in terms of  the components $\psi_n$
in an expansion 
$$
\state{\psi} = \sum_{n=0}^{\infty} \psi_n (\crea)^n \vac.
$$
Let us first take a look at the potential (\ref{Toyaction}) in components:
\be
V(\psi) = \half \sum_n n! (n-1) \psi_n^2 + {1 \over 3} \sum_{m,n,p}
m! n! p!\ G_{mnp} \psi_m \psi_n \psi_p
\label{toypotcomp}
\ee
where the coefficients $G_{mnp}$ are generated by the function:
\bea
G(z_1,z_2,z_3) &=& \exp ( \half \sum_{i,j=1}^3  z_i N_{ij} z_j )\nonumber\\
&\equiv& \sum_{mnp} G_{mnp} (z_1)^m (z_2)^n(z_3)^p.\nonumber
\eea
Due to the form of the matrix $N$, the $G_{mnp}$ are completely symmetric
and are zero when the sum $(m+n+p)$ is odd.
This last property guarantees that the potential  possesses a $\IZ_2$
{\em twist symmetry} just as in the full string field theory.
This symmetry acts on the components as $\psi_n \to (-1)^{n} \psi_n$.
As in the full string field theory, the components that are odd
under the twist symmetry  can be consistently put to zero:
$$
\psi_{2n+1} =0.
$$
The equation (\ref{Toyeom}) for the even components becomes
\be
(2m - 1) \psi_{2m} + \sum_{n,p =0}^\infty (2n)! (2p)! G_{2m,2n,2p} \psi_{2n}
 \psi_{2p} =0.
\label{eomcomp}
\ee
The trivial solution, $\psi_{2m} = 0$, has $V(\psi) =0$ and
 is the one that will correspond to
the unstable state. The solution we are looking for will have lower energy
and will correspond to a local minimum of the potential.

We can also rewrite  the equation of motion (\ref{Toyeom}) as a 
differential equation. Let us use a
short hand notation for the \sfi\ $\psi$: 
$$
\state{\psi} = \sum_{n=0}^{\infty} \psi_n (\crea)^n \vac
\equiv \psi(\crea)\vac.
$$
If we use $ \partial_i = \partial/\partial x_i$, the equation of motion reads
\bea
&&\left( x \frac{\partial}{\partial x} -1 \right) \psi(x)+ \nonumber \\
&&\exp\left( \frac{1}{2} \sum_{i,j = 2}^3 N_{ij} 
\partial_i\partial_j+
x \sum_{i=2}^3 N_{1i} \partial_i+ 
\frac{1}{2} N_{11} x^2 \right)
\left.\psi(x_2) \psi(x_3)\right|_{x_2 = x_3 = 0} = 0. 
\label{diffeom}
\eea
Here we have used that
$$
\leftvac a\ F(\crea) \vac = \left.\frac{\partial}{\partial \crea} 
F(\crea)\right|_{\crea = 0}. 
$$
The resulting equation is a non-linear differential equation of infinite order.

\section{Associativity}\label{TM:ass}
In the full string field theory  the star product is \emph{associative}.
We will now check the associativity in our model on a basis of coherent states. The
star-product of two coherent states is easy to calculate:
$$
e^{l_1 \crea} \vac \ast e^{l_2 \crea} \vac =
A \exp \left( \lambda a^{\dagger 2} + \mu ( l_1 + l_2) \crea\right) \vac
$$
with 
$ A = \exp\left( \lambda (l_1^2 + l_2^2) + \mu\ l_1 l_2 \right)$. Then using
the correlator
\be\label{corr2}
 \leftvac e^{k a^2 + \rho a} e^{l a^{\dagger 2} + \sigma \crea} \vac=
\frac{1}{\sqrt{1 - 4 k l }} 
\exp\left( \frac{l \rho^2 + \sigma \rho + k \sigma^2}{1 - 4 k l }\right),
\ee
we  find
\bq\label{Toy3star}
&&\left(e^{l_1 \crea} \vac \ast e^{l_2 \crea} \vac\right)
\ast e^{l_3 \crea}\vac =\\
&&
A B \frac{1}{\sqrt{1 - 4 \lambda^2}}
\exp\left\{\frac{\lambda \mu^2 (l_1+l_2)^2 + \mu^2 (\crea + l_3) (l_1+l_2) 
+ \lambda (\crea + l_3)^2 \mu^2}{1-4 \lambda^2}\right\} \vac \nonumber
\eq  
with
$B = \exp \left( \lambda a^{\dagger 2} + \lambda l_3^2 + \mu \crea l_3\right)$.
Imposing cyclicity among $l_1,l_2,l_3$ we find that the star-product is
associative only in the following three cases, hereafter called case I, II and
III:
\begin{enumerate}
\item[I.]
$\mu = 0 $, then $N = \left(\begin{array}{ccc}
2 \lambda &0 &0\\
0&2 \lambda & 0\\
0& 0 &2 \lambda\\
\end{array} \right) $
\item[II.] 
$ 2 \lambda = \mu -1$, then $N = \left(\begin{array}{ccc}
\mu-1 &\mu &\mu\\
\mu&\mu-1 & \mu\\
\mu&\mu  &\mu-1\\
\end{array} \right) $
\item[III.] 
$ \lambda = 1/2$, then $N = \left(\begin{array}{ccc}
1 &\mu &\mu\\
\mu&1 & \mu\\
\mu&\mu  &1\\
\end{array} \right) $
\end{enumerate} 
However, due to the factor $1 / \sqrt{1 - 4 \lambda^2}$ in
equation\refpj{Toy3star} the star product of 3 coherent states diverges in the
last case. Therefore we should look for another proof of associativity in 
this case. We will not do this, we just discard this case. 
\section{Derivation of the star-algebra}\label{TM:der}
Let us now look if $D = a - a^\dagger$ is a derivation of the $\ast$-algebra:
$$
D(A \ast B) =DA \ast B + A \ast DB
\mbox{\quad where $A$ and $B$ are two string fields}.
$$
This is analogous to $\alpha^\mu_1-\alpha^\mu_{-1}$ being a derivation 
in the full \sft, see for example~\cite{conserv}.
It is easy to see that for $D$ to be a derivation we need
\be\label{Toyder1}
\sum_i ( a_i - \crea_i)\state{V}=0.
\ee
Let us calculate the left hand side of\refpj{Toyder1}:
\bqs
\sum_i ( a_i - \crea_i)\state{V} &=& 
\sum_i ( \parder{\crea_i} - \crea_i)\state{V}\\
&=& \sum_i ( N_{ij}\crea_j- \crea_i)\state{V}
\eqs
This is zero if and only if $(\ 1\ 1\ 1\ ) \cdot (N-1) = 0$.
\begin{itemize}
\item{case I}\\
We need $3 ( 2 \lambda -1) = 0 $ so $\lambda = 1/2$. Hence $D$ is a derivation
if and only if $N=1$. We will  call this trivial case henceforth case  Id.
\item{case II}\\
We need $ 2 \mu + \mu -2=0$, hence $\mu = 2/3$. In this case we have
\be\label{NIId}
N = \left(\begin{array}{ccc}
-1/3& 2/3&2/3\\
2/3 &-1/3& 2/3\\
2/3 & 2/3& -1/3\\
\end{array} \right)
\ee
Henceforth, we call this subcase IId.
\item{case III}\\
We need $2 \mu =0$ so $\mu=0$. This reduces to case Id.
\end{itemize}

For the case Id we will show in
section \ref{TM:CaseI} that there is no non-perturbative vacuum. Therefore we
consider the value\refpj{NIId} as the most important special case that we
would like to solve exactly in our toy-model.
\paragraph{}
In the full string field theory, an important role is played by the
 {\em identity string field} $I$ i.e.~a
string field obeying $I \ast A = A = A \ast I$ for (almost) all string fields
$A$\footnote{There are some anomalies in the ghost sector, 
$I$ is not an identity of the star algebra on \emph{all} states, see~\cite{conserv}.}.
In the toy model, there exists an identity string field $I$ only in case II.
In this case we have for the identity $I$
$$
\state{I} = \frac{\sqrt{2 \mu -1}}{\mu} 
\exp \left( \frac{1-\mu}{4 \mu-2}a^{\dagger 2} \right) \vac
.$$
Using the correlator\refpj{corr2}, the reader can
easily check that 
$$ \state{I} \ast e^{l \crea} \vac =  e^{l \crea} \vac $$
for all coherent states $e^{l \crea} \vac $, thus proving that $I$ is the
identity. Proving that there is no identity if $N$ does not belong to case II
is most easily done by first arguing that the identity should be a Gaussian in
the creation operator $\crea$ and then showing that one can not find a
Gaussian which acts as the identity on all coherent states. In case IId 
the identity string field reduces to
\be\label{Toyspid}
\state{I} = \frac{2}{\sqrt{3}}\exp \left( \frac{1}{2} a^{\dagger 2} \right) 
\vac
\ee

\section{Exact results in case I}\label{TM:CaseI}
\subsection{Closed form expression for the stable vacuum}
We now construct the exact solution in the case I, where
$$N = \left(\begin{array}{ccc}
2 \lambda &0 &0\\
0&2 \lambda & 0\\
0& 0 &2 \lambda\\
\end{array} \right).
$$
The coefficients $G_{2m,2n,2p}$ entering in the equation of motion
(\ref{eomcomp}) are particularly simple in this case:
$$
G_{2m,2n,2p} = {\l^{m+n+p} \over m! n! p!}.
$$
Equation (\ref{eomcomp}) reduces to
\be
\psi_{2m} =  {\l^m \over (1-2m) m!} g(\l)^2
\label{eomcase1}
\ee
where we have defined a function $g(\l)$ by
$$
g(\l) = \sum_{n=0}^\infty {(2n)! \l^n \over n!} \psi_{2n}(\l).
$$
Multiplying equation (\ref{eomcase1}) by $(2m)! \l^m / m!$
and summing over $m$ we obtain $g(\l)$:
$$
g(\l) = \left( \sum_n {\l^{2n} (2n)! \over (n!)^2 (1 - 2n)} \right)^{-1}
= {1 \over \sqrt{1 - 4 \l^2}}
$$
Hence our candidate for the {\em stable vacuum} $|{\rm vac}\rangle$ is
$$
|{\rm vac}\rangle = {1 \over 1 - 4 \l^2} \sum_{n=0}^\infty 
{\l^n \over  n! (1- 2 n)}
(a^\dagger)^{2n} |0\rangle
$$

Using the representation (\ref{diffeom}), it is also possible to
derive a generating function for the coefficients $\psi_{2n}$. 
Putting $\l = - l^2$, the differential equation (\ref{diffeom}) reduces to
$$
\left(x \parder{x} -1\right) \psi(x) + 
\left.\exp-l^2 \left(\partial_2^2 + \partial_3^2 + x^2\right)\psi(x_2) \psi(x_3)
\right|_{x_2=x_3=0} = 0 
$$
This is
 $$
\left(x \parder{x} -1\right) \psi(x) + e^{ - l ^2 x^2} c^2 = 0, 
$$
where $c$ is just a number 
$$c = \left.e^{- l^2 \partial^2_x} \psi(x) \right|_{x=0}. 
$$
A solution of this differential equation is
$$ \psi(x) = \frac{1}{1-4 l^4} \phi( l x), $$ 
where $\phi(x)$ is the function
\bqs
\phi(x)&=& \exp(-x^2) + \sqrt{\pi} x\ \erf ( x)\\
&=& - \sum_{m=0}^{+\infty}\frac{(- x^2)^m}{m!\ ( 2 m -1)}.
\eqs

The energy difference between the false and true vacuum
 can be expressed entirely
in terms of the function $g(\l)$: 
$$ V({\rm vac}) = - { 1\over 6}
g(\l)^3= - {1 \over 6} (1 -4 \l^2)^{-3/2}. $$
 It is
clear that the true vacuum only exists for $|\l| < \half$ since
the value of the potential becomes imaginary outside this range.
Also, for $|\l| > \half$, the state $|{\rm vac}\rangle$ is no
longer normalisable. Note that for the special
case Id, $\l = \half$, there does not seem to be a true vacuum.

\subsection{Closed form expression for the effective potential}
We can also determine  the exact {\em effective tachyon potential} $V(t)$
 by solving
for the $\psi_{2n},\ n>~0$ in terms of $t \equiv \psi_0$. The
equation for these components becomes:
\be \psi_{2m}(\l,t)  =  {\l^m
\over (1-2m) m!} ( t + h(\l,t))^2 \qquad {\rm for}\ m>0
\label{tachpoteom}
\ee
where we have defined
$$ h(\l, t) =
\sum_{n=1}^\infty    {(2n)! \l^n \over n!} \psi_{2n}(\l,t).
 $$
Multiplying equation (\ref{tachpoteom}) by $(2m)! \l^m / m!$ and
summing over $m$ we get a quadratic equation for $h(\l, t)$:
$$
h(\l, t) = (\sqrt{ 1- 4 \l^2} - 1) (t + h(\l, t))^2 .$$
The two
solutions $h_\pm$
$$ h_\pm = {1 \over 2 (1- \sqrt{1- 4
\l^2})}\left( - 2 t( 1- \sqrt{1 - 4 \l^2}) -1 \pm \sqrt{4t( 1-
\sqrt{1 - 4 \l^2}) + 1} \right) $$
will give rise to two branches
of the effective potential. When we also impose the equation for
$t$, we see that the unstable vacuum $t=0$ and the stable vacuum
$t = {1 \over 1 - 4 \l^2}$ lie on the same branch (i.e. the one
determined by $h_+$) just as in the full string field theory. Substituting
$h_\pm$ in (\ref{tachpoteom}) to obtain the coefficients
$\psi_{2n\pm}(\l,t)$ and substituting those in (\ref{toypotcomp}) we
find the exact form of the two branches of the effective
potential $V_\pm (t)$:
$$ V_\pm =  - \half t^2 +  {h_\pm^2
\over 2( 1 - \sqrt{1 - 4 \l^2})} + {1 \over 3} (t + h_\pm)^3.
 $$
 As is
the case in the full bosonic string field theory, the
branch $V_+(t)$, which links the unstable and the stable vacuum, terminates
at a finite negative value $t_*$, given in this case by 
\be 
t_* = - {1 \over 4(  1- \sqrt{1 - 4 \l^2})}.
\label{tstar1}
\ee
 At this point, the
two branches meet. It is also the only point where they intersect,
since $V_- > V_+ $ for all other values of $t$.

\subsection{The level truncation method}
We  can also discuss the convergence of the level truncation
method in this case. We will focus on the level $(2k,6k)$
approximation to the tachyon potential. This means that we
include the fields up to level $2k$ and keep all the terms in the
potential involving these fields. 
In this approximation, the equation for the extremum
is just  (\ref{eomcomp}) with all sums now running from $0$ to
$k$. The solution proceeds just as in the previous section.
First one solves for the function $g^{(k)}(\l)$: 
$$ g^{(k)}(\l) \equiv
\left( \sum_{n=0}^k {\l^{2n} (2n)! \over (n!)^2 (1 - 2n)}
 \right)^{-1}
= \left(\sqrt{1 - 4 \l^2} + E (\l, k)\right)^{-1}.
$$
The function $E (\l, k)$, which represents the error we make by
truncating at level $2k$, can be 
 expressed in terms of special functions
$$
E (\l, k)= { 2^{1 + 2 k} \l^{2(1 + k)} \G (\half + k )\ _2 F_1 ( 1, \half
+ k, 2 + k; 4 \l^2) \over \sqrt{\pi} (k+1)!}.
$$
The level-truncated expressions for the  components of the
 approximate vacuum state $|{\rm vac}^{(k)}\rangle$
 and the value of $V_{(2k, 6k)}$ at the minimum  are
given by:
\bea
\psi^{(k)}_{2m} &=& {\l^m \over (1-2m) m!} g^{(k)}(\l)^2\nonumber \\
V_{(2k, 6k)}( {\rm vac}^{(k)} ) &=&  - {1 \over 6 }g^{(k)}(\l)^3
\nonumber
\eea

\begin{figure}
\begin{center}
\begin{psfrags}
\psfrag{V}[][]{$V(t)$}
\psfrag{t}[][]{$t$}
\psfrag{a}[][]{{\scriptsize$(0,0)$}}
\psfrag{b}[][]{{\scriptsize$(2,6)$}}
\psfrag{c}[][]{{\scriptsize$(4,12)$}}
\psfrag{d}[][]{{\scriptsize$(6,18)$}}
\psfrag{e}[][]{{\scriptsize exact}}
\epsfig{file=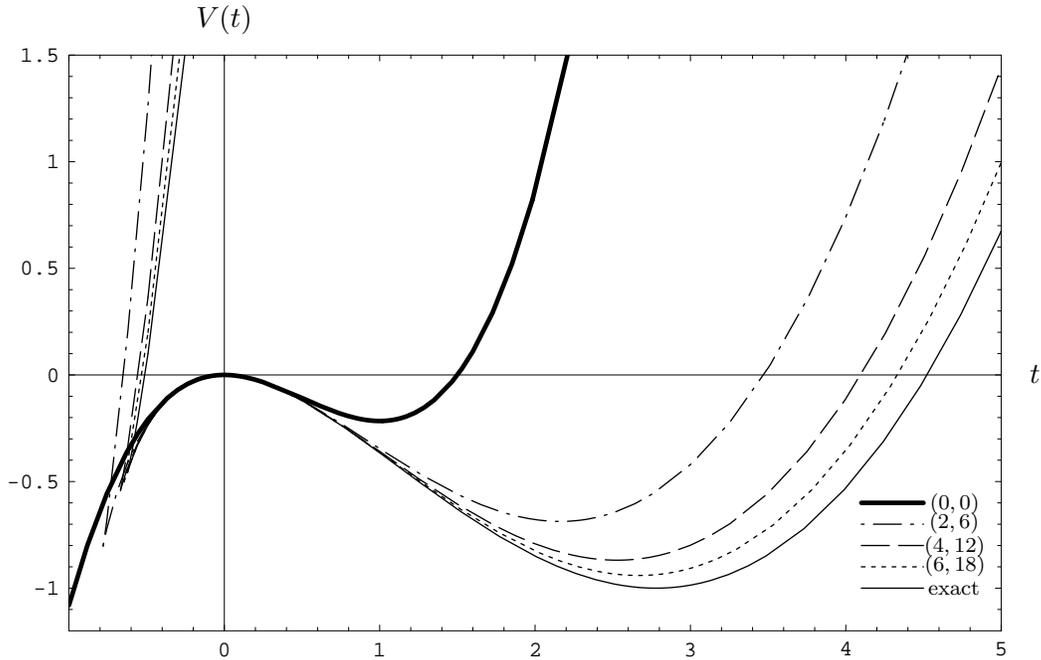, width=400pt}
\end{psfrags}
\end{center}
\caption{The level-truncated effective potential 
 for $\l = 0.4$
at level $(0,0)$, level $(2,6)$, level $(4,12)$ and level $(6,18)$ 
as compared to the exact result. We have rescaled the potential by a factor
$6 ( 1 - 4 \l^2)^{3/2}$ so that the minimum occurs at $V = -1$.
\label{toytrunc}}
\end{figure}

The determination of the level-truncated effective tachyon potential
also proceeds as before.
The result is
$$ V_{(2k,6k)\pm}(t) = - \half t^2 + { {h^{(k)}_\pm}^2
\over 2( 1 - \sqrt{1 - 4 \l^2}- E(\l,k))} + {1 \over 3} (t + h^{(k)}_\pm)^3
 $$
 with
\bea 
 h^{(k)}_\pm   &=& {1 \over 2 (1- \sqrt{1- 4
\l^2}- E(\l,k))}\Big( - 2 t\big( 1- \sqrt{1 - 4 \l^2} - E(\l,k)\big)
-1 \nonumber \\ 
&\ &\pm \sqrt{4t( 1-
\sqrt{1 - 4 \l^2}- E(\l,k)) + 1} \Big).\nonumber
\eea
 Again, the potential has two branches which 
intersect at a finite negative value $t_*^{(k)}$:
\be
t_*^{(k)} = - {1 \over 4 ( 1 - \sqrt{1 - 4 \l^2} - E(\l, k))}
\label{tstar2}
\ee 
A plot of both branches of the potential for $k= 0,\ 1,\ 2$ at $\l = 0.4$,
as compared to the exact result, is shown in figure 
\ref{toytrunc}.



\subsection{Convergence properties and comparison to the full string field 
theory}
The results of the previous sections allow us to derive some exact results
concerning the convergence properties of the level truncation
method in this model and to compare them with the behaviour found in
the full string field theory using numerical methods \cite{mt}. 
For this purpose, we need the asymptotic behaviour of the function $E(\l, k)$
 for large level
$k$ \cite{abramowitz}:
\be
E (\l,k) \sim {2 \l^2 \over \sqrt{\pi} (1 - 4 \l^2)} k^{-3/2} (4 \l^2)^k
[ 1 + {\cal O} (k^{-1})] \qquad {\rm for}\ k \to \infty.
\label{asympt}
\ee
Hence
the error we make in the level approximation to the coefficients of the true
vacuum and the value of the potential at its minimum
  goes like
\bea
\psi_{2m}-\psi^{(k)}_{2m} &\sim& { 2^{2k+2} \l^{2k+m+2}k^{-3/2}
\over \sqrt{\pi} m! (1 - 2m) ( 1 - 4 \l^2)^{5/2}}\nonumber \\
V( {\rm vac} )- V_{(2k, 6k)}( {\rm vac}^{(k)} )&\sim&
-{ \l^2 k^{-3/2} (4 \l^2)^k \over \sqrt{\pi} (1 - 4 \l^2)^{3}}\nonumber
\eea
for large level $k$. 
We see that, both for the components
of the vacuum state and the value of the potential at the minimum,
the level truncation method  converges
to the exact answer in a manner which is essentially exponential as 
a function of the level: it goes
like $k^{-3/2} e^{- k |\ln 4 \l^2|}$.
This exponential behaviour is comparable
 to the one found `experimentally' in the full string field theory 
problem in \cite{mt}: there, the
error was found to behave like $({1 \over 3})^k$.  

The effective tachyon potential
in the toy model has a finite radius of convergence $|t_*|$ as in 
the full string field theory. In the level truncation method, the
radius of convergence $|t_*^{(k)}|$ rapidly approaches the exact value;
indeed, from (\ref{tstar1}), (\ref{tstar2}) and (\ref{asympt})  we have
$$
t_* -  t_*^{(k)} \sim {\l^2 \over 8\sqrt{\pi}
 ( 1- \sqrt{1 - 4 \l^2})^2 ( 1- 4 \l^2)} k^{-3/2} (4 \l^2 )^k \qquad 
 {\rm for}\ k \to \infty.\
$$
In contrast to the string field theory effective potential \cite{kost, mt},
the toy model effective potential does not display a breakdown of convergence
for positive values of $t$.


\section{Other exact solutions}\label{TM:othersols}
We can also find the exact minimum in case II when $\mu = 1$, i.e.~when 
$$N = \left(\begin{array}{ccc}
0&1 &1\\
1&0 &1\\
1&1 &0\\
\end{array} \right). $$
We need to solve the following equation:
$$
\left(x \parder{x} -1\right) \psi(x) + 
\left.\exp\left(\partial_2 \partial_3  + x( \partial_2 + \partial_3)\right)\psi(x_2) \psi(x_3)
\right|_{x_2=x_3=0} = 0 
.$$
A solution of this equation is $\psi(x) = 1$. 
$$
\state{\mbox{false vac}} = 0 \vac \qquad
\state{\mbox{true vac}} = 1 \vac
$$
More generally we can also solve $$N = \left(\begin{array}{ccc}
0&\mu &\mu\\
\mu&0 &\mu\\
\mu&\mu &0\\
\end{array} \right),$$
for general $\mu$, again the solution is  $\psi(x) = 1$. However this case is
not associative if $\mu\neq 1$.
\section{Towards the exact solution in case IId?}\label{TM:CaseIId}
\subsection{The star product in momentum space}
In section~\ref{TM:ass} we have deduced that $D = a-\crea$ is a
derivation of the star algebra. If we write the creation and 
annihilation operators in terms of the momentum 
and coordinate operators:
$$
\left\{
\begin{array}{ll}
\crea=& \frac{1}{\sqrt{2}}\ (p + i x),\\
a=& \frac{1}{\sqrt{2}}\ (p -i x),\\
\end{array}\right.
$$
we see that $D$ is proportional to $\p/\p p$. 
Therefore it is tempting to anticipate that the star product will reduce
to an ordinary product in momentum space, and this is indeed the case\footnote{In
Witten's \sft\ the operators $D_n^{\mu} = \alpha^{\mu}_n + (-1)^n
\alpha^{\mu}_{-n}$ are derivations of the star algebra. This suggests going to
the $k$ -- space for the odd matter oscillators $\alpha^{\mu}_{2 n +1}$ and to
the $x$ -- space for the even matter oscillators $\alpha^{\mu}_{2 n}$.
See~\cite{Bars} where an analysis along these lines was performed. In Witten's
string field theory the star product reduces to a matrix product in the split
string formalism~\cite{split}.}. If we write the
states in momentum representation:
$$\state{\psi} = \int dp\  \psi(p) \state{p}_p,$$
where the states $\state{p}_p$ are the eigenstates of the momentum operator
$\hat p$, normalized in such a way that $\langle p_1 \state{p_2} =
\delta(p_1-p_2)$ --- we use the extra subscript to denote which representation we
are using. We find 
\be\label{Toyordprod}
\state{\psi} \ast \state{\eta} = \int dp\ \pi^{1/4} \sqrt{\frac{3}{2}}\ \psi(p)
\eta(p) \state{p}_p
\ee
and
\be\label{Toybew3}
\etats{V} \state{\psi_1} \state{\psi_2}\state{\psi_3} = \pi^{1/4}
\sqrt{\frac{3}{2}} \int dp\  \psi_1(p) \psi_2(p)  \psi_3(p).
\ee
This last equation is easy to prove on a basis of coherent states. 
If $\state{\psi_i} = \exp\ ( l_i \crea )\vac$, then
$$ \etats{V} \state{\psi_1}\state{\psi_2} \state{\psi_3} = e^{\ l^T N l}.$$
Let us verify if we get the same result in momentum space. A coherent state is 
given by a Gaussian in momentum space:
$$ e^{ l \crea} \vac_a =  
\frac{1}{\pi^{1/4}} \exp(-\frac{1}{2}\ l^2+ \sqrt{2}\ l\ k -\frac{k^2}{2}) 
.$$
Equation\refpj{Toybew3} then holds by Gaussian integration.

As a check on our result we will verify that the state $\state{I}$ given
by\refpj{Toyspid} is the identity in momentum space. In momentum space we have
$$
\state{I} = \sqrt{\frac{2}{3}}\ \frac{1}{\pi^{1/4}}\  1 \mbox{\qquad as a function in
momentum space}, $$
therefore we have for arbitrary states $\psi$
$$ \state{I} \ast \state{\psi} = \sqrt{\frac{2}{3}}\ \frac{1}{\pi^{1/4}}\  1
 \cdot
	\pi^{1/4}\sqrt{\frac{3}{2}}\  \psi(k) = \psi(k), $$
as it should be.       	    
\subsection{The equation of motion in momentum space}
The equation of motion we want to solve now becomes in momentum space
\bqs
&&(a^\dagger a -1) \state{\psi} +\state{\psi} \ast \state{\psi} \\
&& = {1 \over 2} \left( - {\p^2 \over \p p^2} + (p^2 - 3)
 \right)\psi + \pi^{1/4}
\sqrt{{3 \over 2}}\psi(p)^2 =0.  
\eqs
If we drop some constants, the differential equation we are left with reads
\be
{\p^2 \over \p p^2} \psi(p) = (p^2 - 3) \psi(p) + \psi(p)^2.
\label{oureq}
\ee
So we see that instead of the infinite order differential equation we started
with, we have now a second order non-linear differential equation.
A large body of literature exists (see e.g. \cite{kamke}) 
on second order differential equations
that have the Painlev\'{e} property, meaning that the solutions to these
equations have no movable critical points. Such equations can be transformed
to one of 50 equations whose solutions can be expressed in terms
of known transcendental functions.  Applying the algorithm described 
in \cite{ablowitz}, one finds that equation\refpj{oureq} is not of
the Painlev\'{e} type due to the presence of movable
logarithmic singularities. Hence we have been unsuccesful in 
solving\refpj{oureq}.

\subsection{Numerical results}
Even though we are not able to find a closed form solution in this case, 
we can get good approximate results with the level truncation method. 
We give the potential including fields up to level 4. It reads:
\bqs
&&V(\state{\psi})=\frac{-{\psi_0}^2}{2} + 
  \frac{{\psi_0}^3}{3} - 
  \frac{{\psi_0}^2\,\psi_2}{3} + 
  {\psi_2}^2 + 
  \psi_0\,{\psi_2}^2 + 
  \frac{13\,{\psi_2}^3}{27} + 
  \frac{{\psi_0}^2\,\psi_4}{3}\\&& - 
  \frac{34\,\psi_0\,\psi_2\,
     \psi_4}{9} + 
  \frac{41\,{\psi_2}^2\,\psi_4}
   {27} + 36\,{\psi_4}^2 + 
  \frac{227\,\psi_0\,{\psi_4}^2}
   {27} + \frac{319\,\psi_2\,
     {\psi_4}^2}{27} + 
  \frac{1249\,{\psi_4}^3}{81}
\eqs
We can minimize this action and we find
\begin{description}
\item{at level $0$:}  $ \state{\psi} = 1.\vac$\\ with $V(\psi) =
-0.166667$.
\item{at level $2$:}  $\state{\psi} = (1.05083\ +0.0870701\ a^{\dagger 2})\vac$
\\
with $V(\psi) = -0.181514$
\item{at level $4$:}  $\state{\psi} = (1.0508\  +0.0867394\ a^{\dagger 2}
  -0.000383389\ a^{\dagger 4})\vac$ \\
with $V(\psi) = -0.181521$
\item{at level $6$:}  $\state{\psi} = (1.05082\  +0.0867768\ a^{\dagger 2}
  -0.000408059\ a^{\dagger 4}-0.0000352206\  a^{\dagger 6})\vac$\\ 
with $V(\psi) = -0.181523$
\item{at level $8$:}  $
\state{\psi} = (1.05082\  +0.0867771\ a^{\dagger 2}
  -0.000412528\ a^{\dagger 4}-0.0000341415\  a^{\dagger 6}$\\
  \hspace*{2.5 cm}$+1.788 \cdot 10^{-6}\ a^{\dagger 8})\vac $\\
with $V(\psi) =-0.181524 $
\item{at level $10$:}  $
\state{\psi} = (1.05082\  +0.0867771\ a^{\dagger 2}
  -0.000412537\ a^{\dagger 4}-0.0000339848\  a^{\dagger 6}$\\
  \hspace*{2.5 cm}$+1.76475 \cdot 10^{-6} a^{\dagger 8}
  -4.54233\cdot 10^{-8}a^{\dagger 10})\vac $\\
with $V(\psi) = -0.181524$
\end{description}
 We see  that the level truncation
method clearly converges to some definite answer.
\section{Conclusions and topics for further research}\label{TM:end}
We simplified Witten's open \sft\ by dropping all the ghosts and keeping only
one matter oscillator. The model we constructed closely resembles the full \sft\
on the following points:
\begin{itemize}
\item
There is a false vacuum and a stable vacuum.
\item
The interaction is given in terms of ``Neumann coefficients'' and can be 
written
by using an associative star product.
\item
There is a notion of level truncation which converges rapidly to the correct
answer.
\end{itemize}
For some special values of one of the parameters of the
model, we were able to obtain the exact solution for the stable 
vacuum state and
the value of the potential at the minimum. 

For other values of the parameters we did not succeed in 
constructing the exact minimum of the tachyon potential. 
This does not mean that it is impossible to solve 
Witten's \sft\ exactly. In the full \sft\ there is a lot
more symmetry around: for example Witten's \sft\ has a huge gauge invariance
and one could try to solve the equation of motion by 
making a pure -- gauge like
ansatz~\cite{Schnabl}.

Therefore maybe a natural thing to
do is to set up a toy model that includes some of the ghost oscillators in such
a way that there is also a gauge invariance. 
Another research topic would be to set up a toy model of Berkovits' 
superstring field theory (see \cite{berk} for a recent review). It also
should not be too difficult to try to mathematically prove the convergence 
of the level truncation method
 in these toy models. This might teach us something
about why the level truncation method converges in the full 
string field theory.

\acknowledgments{This work was supported in part
by the European Commission RTN project HPRN-CT-2000-00131. 
The authors would like to thank Walter Troost for discussions and especially
Martin Schnabl for collaboration on several parts of this work. 
P.J.D.S. is aspirant FWO-Vlaanderen.}

\end{document}